\documentclass[preprint, prX]{revtex4}

\usepackage{amsmath}    
\usepackage{graphicx}   
\usepackage{verbatim}   
\usepackage{color}      
\usepackage{subfigure}  
\usepackage{hyperref}   
\usepackage{amssymb}    
\usepackage{epsfig}
\usepackage{graphics,graphicx}
\usepackage{setspace}
\usepackage{url}
\usepackage{algorithm,algorithmic}

\begin{document}

\begin{abstract}
Measuring time in mass sports competitions is usually performed using expensive measuring devices. Unfortunately, these solutions are not acceptable by many organizers of sporting competitions. In order to make the measuring time as cheap as possible, the domain-specific language (DSL) EasyTime was proposed. In practice, it has been proven to be universal, flexible, and efficient. It can even reduce the number of required measuring devices. On the other hand, programming in EasyTime is not easy, because it requires a domain-expert to program in a textual manner. In this paper, the domain-specific modeling language (DSML) EasyTime II is proposed, which simplifies the programming of the measuring system. First, the DSL EasyTime domain analysis is presented. Then, the development of DSML is described in detail. Finally, the DSML was tested by regular organizers of a sporting competition. This test showed that DSML can be used by end-users without any previous programming knowledge.

\textit{To cite paper as follows: I. Fister Jr., T. Kosar, M. Mernik, I. Fister, Upgrading EasyTime: from a textual to a visual language, In {\em Proceedings of the 21st International Electrotechnical and Computer Science Conference}, Portoro\v{z}, Slovenia, 2012.
}

\end{abstract}

\title{Upgrading EasyTime: from a textual to a visual language}

\author{Iztok Fister Jr.}
\altaffiliation{University of Maribor, Faculty of electrical engineering and computer science
Smetanova 17, 2000 Maribor}
\email{iztok.fister@guest.arnes.si}

\author{Toma\v{z} Kosar}
\altaffiliation{University of Maribor, Faculty of electrical engineering and computer science
Smetanova 17, 2000 Maribor}
\email{tomaz.kosar@uni-mb.si}

\author{Marjan Mernik}
\altaffiliation{University of Maribor, Faculty of electrical engineering and computer science
Smetanova 17, 2000 Maribor}
\email{marjan.mernik@uni-mb.si}

\author{Iztok Fister}
\altaffiliation{University of Maribor, Faculty of electrical engineering and computer science
Smetanova 17, 2000 Maribor}
\email{iztok.fister@uni-mb.si}

\maketitle

\section{Introduction}

The problem of measuring time in sporting competitions is relatively old. Many approaches have been developed to deal with this problem. One of the more efficient solutions was the domain-specific language (DSL) \cite{Mernik:2005}\cite{vanDeursen:2000} EasyTime\cite{Fister:2011}. The development of EasyTime arose from the need to cover the results of double triathlon competition in 2009, but outgrown limits of this particular triathlon competition. After its first successful use in practice, two demands for the future development of EasyTime were revealed:
\begin{itemize}
\item how to satisfy the demands of various competitions,
\item how to simplify the handling of the measuring system in order for it also be usable for regular organizers of sporting competitions.
\end{itemize}
The first demand was satisfied with EasyTime. EasyTime is a small and efficient DSL with high expressive-power. Until now, it has been applied to various sports competitions like triathlons, time-trials in bicycling, cyclo-cross, running, etc. Unfortunately, its widespread usage was limited because of a lack of measuring devices. Specifically, the equipment for measuring times in a swim course is very expensive. Therefore, we concentrated on measuring time in relatively small competitions. It is encouraging that many industries showed an interest in using EasyTime for their measuring systems. 

Although the original EasyTime demonstrated itself to be robust and universal in the practice, it still required a domain-specific expert to program the measuring domain. Therefore, the development of a domain-specific modeling language (DSML) EasyTime II was proposed that could satisfy the second demand. Note that DSMLs are currently one of the most interesting research topics in the area of computer languages.  

In this paper, the development of an EasyTime II DSML is presented that consists of four core stages: meta-model construction, the definition of a graphical model, obtaining the semantic model, and code generation. In the beginning, the meta-model for our language is defined. Then, the graphical elements are designed that represent the language concepts and are visible to a user on a panel. In the third stage, the semantic model is obtained. Finally, this model is transformed into an executable code (e.g., Java code) with a model-to-code transformation. 

The structure of this paper is as follows: in Section 2 we describe the problem of measuring time in sports competitions. Section 3 presents DSML EasyTime II and its development stages from the meta-model to code generation. Section 4 describes the practical example using this DSML. The paper concludes with practical experience and directions for future development.

\section{Measuring time in sporting competitions}

Not long ago, measuring time in sporting competitions was performed manually by people who wrote the results for each competitor on paper and at the end, put the competitors in order according to their achievements. This way of measuring time is impossible nowadays because of the large number of participants. On the other hand, there are many modern sports, e.g., the triathlon, aquathlon, etc. that require a very precise measurement of results. Therefore, there is a significant need for electronic measuring devices. These devices need to work precisely and securely in all weather conditions, i.e., in rain and snow. 

"Multi-sport" refers to competitions, where more than one discipline is involved. The most popular multi-sport disciplines are:
\begin{itemize}
\item triathlon (consists of swimming, biking and running),
\item duathlon (consists of running, biking and running),
\item aquathlon (consists of swimming and running),
\item winter triathlon (consists of running, mountain biking, cross-country skiing),
\item etc...
\end{itemize}
Moreover, all these disciplines consist of courses with various distances. For example, a triathlon is divided according to distance into: Ironman, Half Ironman (also Ironman 70.3), Olympic triathlon, Sprint triathlon, etc. Measuring time for these multi-sport competitions is more complicated because of their long duration and a number of the competitors. 

\begin{figure}[htb]
\begin{center}
\includegraphics[scale=0.27]{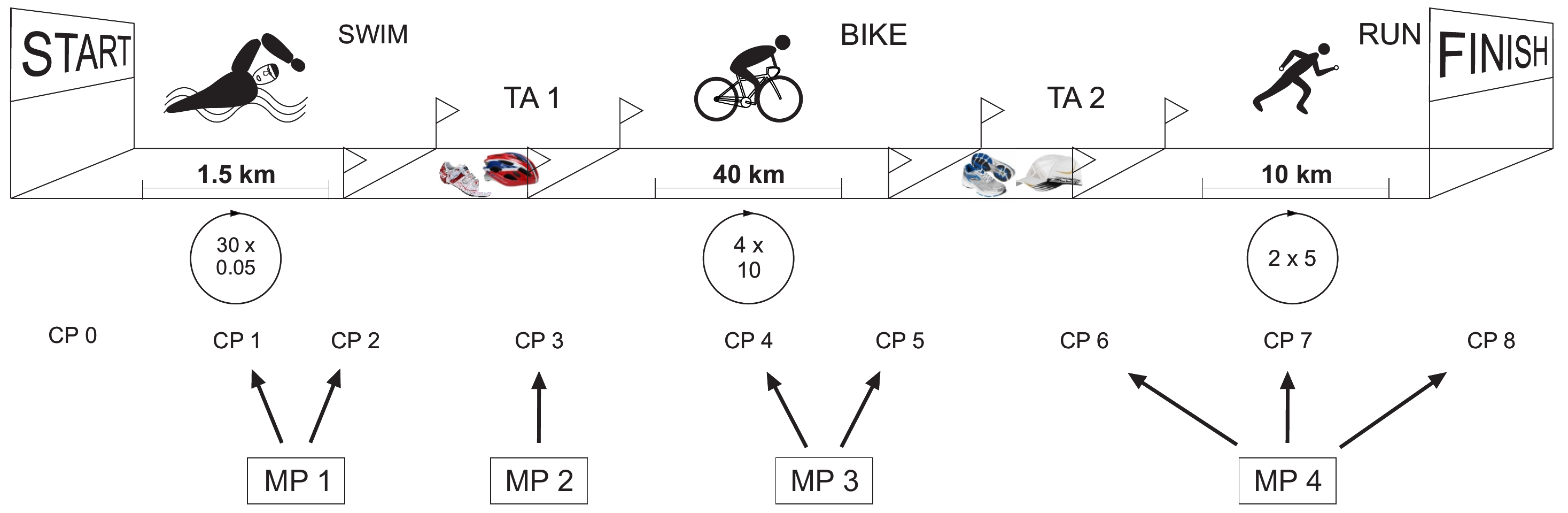} %
\caption{Olympic triathlon}
\label{pic:Olympic}
\end{center}
\vspace{-5mm}
\end{figure}

In Figure~\ref{pic:Olympic}, the Olympic triathlon is presented. The Olympic triathlon consists of 1.5 km of swimming, 40 km of biking and 10 km of running~\cite{IRONMAN-Austria}. Additionally, competitors have to go through two transition areas. In the first transition area, the competitor leaves their swimming equipment and prepares for biking. In the second transition area, the competitor has to drop their bike and prepares for running. Every discipline in this triathlon is split into laps. Therefore, in addition to measuring time, it is also necessary to count laps. 

On the other hand, "single-sport" competitions consist of only one sport (e.g., running, swimming, cycling, etc.). Measuring time in these competitions is not as difficult as it is in multi-sports.

\section{Design and implementation of EasyTime II}

DSMLs have been developed in a number of areas to facilitate the construction of models at a level closer to the conceptual model, thereby making model implementation more accessible to domain experts \cite{ecology}\cite{sprinkle1}\cite{GrayJ}\cite{Czarnecki}. DSMLs are a special kind of languages, where the user does not need to write code. These languages go to the 4th generation of computer languages and are currently one of the most interesting topics of research in the area of computer languages. These languages have a bright future, because of their simple usage. The design and implementation of these languages is a bit more complicated and can be split into the following four core stages: 
\begin{itemize}
\item meta-model construction,
\item definition of a graphical model,
\item obtaining the semantic model, and
\item code generation.
\end{itemize}
The development of DSML EasyTime II grew out of DSL EasyTime. Note that EasyTime is a little textual language for measuring time in sports competitions, which is very efficient and assures the flexibility of a measuring system. The language is based on the compiler/interpreter implementation approach~\cite{Kosar1}. For more information, the design and implementation of DSL EasyTime has been presented in more detail in \cite{Fister:2011}\cite{Fister:2011a}\cite{Fister:2012}. 

However, each DSL development started with a \textit{domain analysis}, in which the \textit{concepts} of DSL are defined and represented within a \textit{feature diagram}. The feature diagram describes the dependencies between these concepts. Furthermore, the concepts can be broken-down into \textit{features} and \textit{sub-features}. Let us remember the main concept of application domain measuring time in triathlon consists of the following features: \textit{events}, \textit{control points}, \textit{measuring time}, \textit{transition areas} and \textit{agents}. Events arise via different disciplines, e.g., sub-features, \textit{swimming}, \textit{cycling}, and \textit{running}. Each control point is described by its \textit{start} and \textit{finish} time together with the number of \textit{laps} to go. The feature \textit{transition area} can be calculated by the difference between the finish and start times, while the \textit{measuring place} is determined by the sub-features \textit{updating time} and \textit{decreasing laps}. Finally, the feature \textit{agent}, which is dedicated to processing events received from the measuring device, can act either \textit{automatically} or \textit{manually}. 

The feature diagram served as a basis for EasyTime II development. Fortunately, this diagram can be incorporated into the Eclipse Modeling Framework tool (EMF) \cite{emf}. The feature diagram was used as a reference for the construction of a \textit{meta-model}. Furthermore, the meta-model served as a basis for the Eclipse Graphical Modeling tool (GMF) \cite{EclipseGmf}\cite{gmf}, in which the used graphical interface (GUI) is defined. Then, the model transformations must be defined in order to call the domain framework, which is a platform that provides functions to implement the semantics of DSMLs in a specific environment. In order to obtain a \textit{semantic model}, this GUI is mapped to the EMF concepts. Finally, this semantic model is translated into an executable code - Java code, which is executed on a Java Virtual Machine. In the rest of paper, the DSML development stages are described in more detail.

\subsection{Meta-model construction}

\textit{Meta-modeling} is the construction of a collection of concepts within a certain domain presented in a context diagram. A model is an abstraction of phenomena in the real world. That is, the meta-model highlights the properties of the real processes. As a result, the model conforms to its meta-model in the way that a computer program conforms to the grammar of the programming language in which it is written~\cite{Tolvanen}\cite{monticore}\cite{GrayJ}. In essence, with a meta-model, a business logic of a process that performs the measuring time in triathlon is described. 

\begin{figure}[htb]
\begin{center}
\includegraphics[scale=0.66]{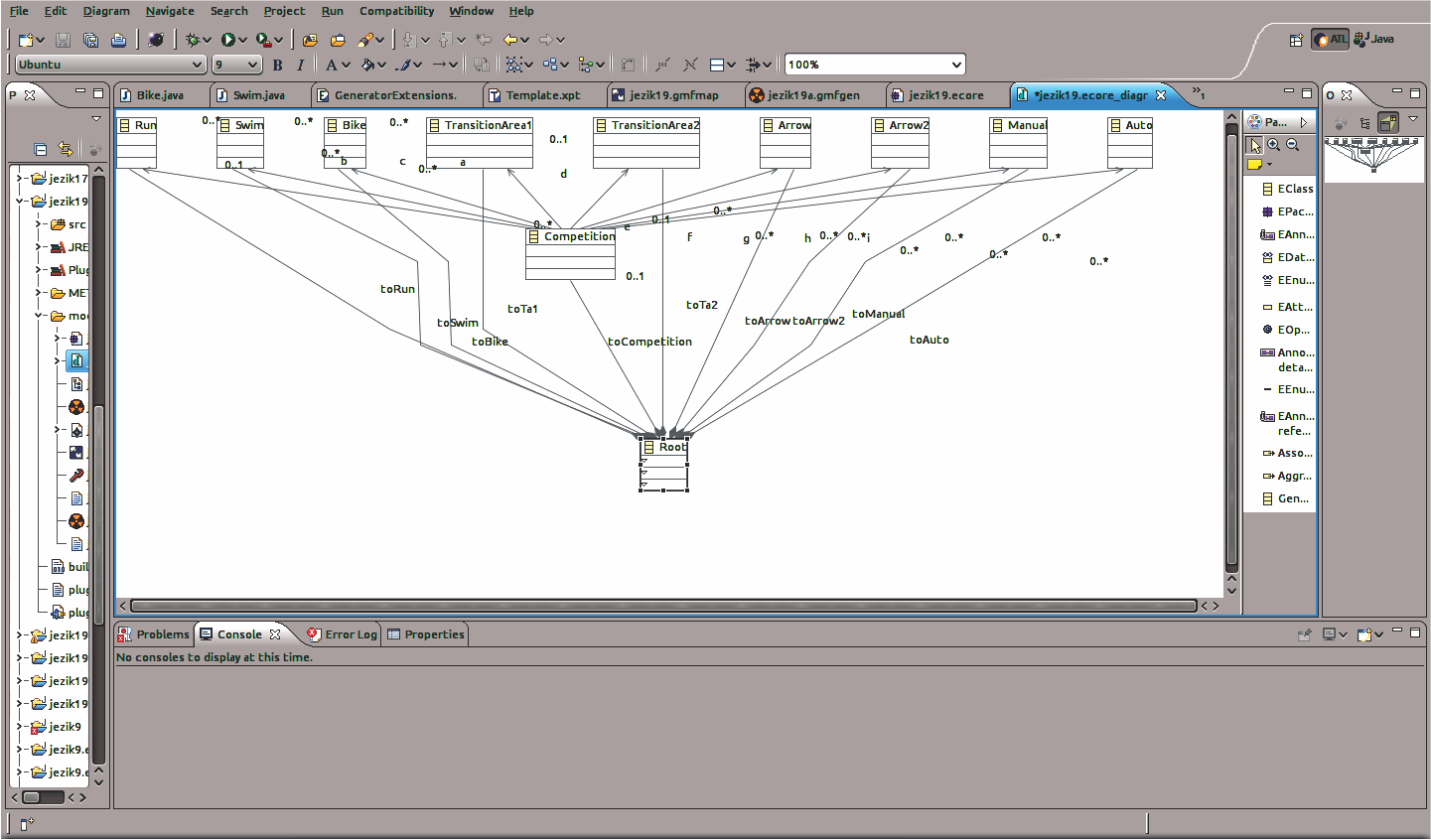} %
\caption{Meta-model example}
\label{pic:meta}
\end{center}
\vspace{-5mm}
\end{figure}

EMF allows us to create good meta-models in a very simple way. The meta-model that performs the triathlon presented in Figure~\ref{pic:Olympic} is presented in Figure~\ref{pic:meta}. The meta-model was developed in Ecore notation. In Figure~\ref{pic:meta}, it can be seen that this meta-model is a conceptual class diagram (defines a set of concepts in the form of classes together with relations).  

This meta-model consists of two main features: root, and competition. These main features are connected
with sub features: swim, bike, run, transition area 1, transition area 2, arrow, arrow2, auto
and manual. These sub-features are linked with features by a connection. Usually, cardinality 0..*
is used. Cardinality 0..* means that we might use this feature exactly 0 or more times in
our model. 

\subsection{Definition of the graphical model}

EasyTime II was modeled using the Eclipse Graphical Modeling Framework (GMF). GMF allow us to create visual aspects of a generated graphical editor \cite{ibm}. These visual aspects consist of the following definitions:
\begin{itemize}
\item graphical definition,
\item tooling definition,
\item mapping definition.
\end{itemize}
In a graphical definition, we choose the elements that will be shown to the user. Then, a tooling definition is performed, in which a visual model is defined, i.e., the palette and toolbox. Finally, the meta-model (business logic) is mapped into a visual model (graphical and tooling definition).

\begin{figure}[htb]
\begin{center}
\includegraphics[scale=1.4]{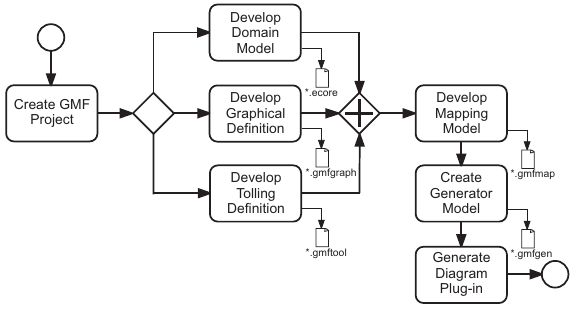} %
\caption{GMF}
\label{pic:gmf}
\end{center}
\vspace{-5mm}
\end{figure}

A sample of the graphical model definition for measuring time in sporting competitions is illustrated in Figure~\ref{pic:gmf}. 

In GMF, images are embedded into a modeling environment. These images are created by GIMP and Inkscape editors, and are shown in the model editor. For tooling definition model, the pictures that are shown in the toolbox are created. 

\subsection{Obtaining the semantic model}

It is not sufficient to complete a DSML definition by only specifying the notions and their representations. The complete definition of DSML requires the semantics of language concepts. Therefore, the abstract syntax defined with Ecore is mapped  into the function calls from the measuring time environment. These mappings lead to model transformations that are applied on EasyTime model instances at runtime in order to obtain their counterparts in real EasyTime infrastructures. The model-to-code transformations can be written in MOFScript, where rules contain calls to the domain framework of the EasyTime system. 

\subsection{Code generation}
Currently, the definition of the model-to-code transformation is still in progress.

\section{Working with EasyTime II}

\begin{figure*}[htb]
\vspace{-5mm}
\begin{center}
\includegraphics[scale=0.36]{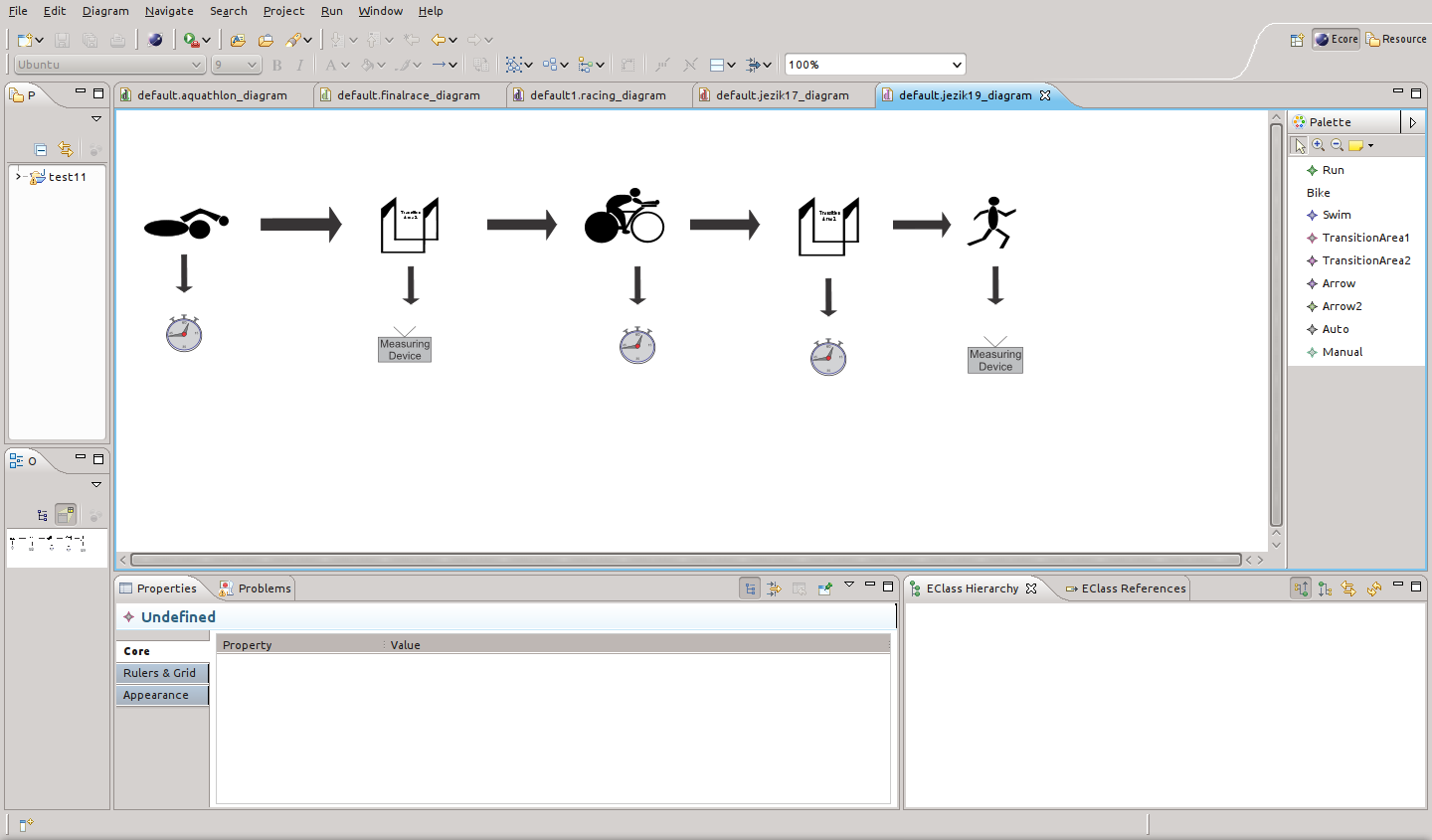} %
\caption{EasyTime II in action}
\label{pic:dsml_example}
\end{center}
\vspace{-5mm}
\end{figure*}

In Figure~\ref{pic:dsml_example}, the measuring time for the Olympic triathlon is presented. Users control the measuring with a simple toolbox, where they can choose between the following sports elements:
\begin{itemize}
\item swim (the graphical representation of swimming is a swimmer),
\item bike (the graphical representation of cycling is a cyclist), 
\item run (the graphical representation of running is a runner).
\end{itemize}
Furthermore, users can select between elements that symbolize:
\begin{itemize}
\item transition area 1, and
\item transition area 2.
\end{itemize}
For agents, users can select between:
\begin{itemize}
\item a manual agent (the graphical representation of the manual agent is a simple clock), and
\item an automatic agent (the graphical representation of the automatic agent is a computer).
\end{itemize}
The arrows symbolize the order in which the sports are performed, and connect the particular sport with the agents.

\section{Conclusion}

In this paper, we presented the design and implementation of DSML EasyTime II. EasyTime II is an extension of DSL EasyTime and was developed to simplify the measuring of time in real sport competitions. In contrast to its predecessor, which required a domain expert to program the measuring system, EasyTime II is dedicated to ordinary users that can control the complicated tasks through easy to use graphical elements. In the future, we intent to measure time in a real-world competition.

\bigskip{\small \smallskip\noindent Updated 20 August 2012.}
\end{document}